# O(a) improvement of the axial current in lattice QCD to one-loop order of perturbation theory

Martin Lüscher
Deutsches Elektronen-Synchrotron DESY
Notkestrasse 85, D-22603 Hamburg, Germany

Peter Weisz
Max-Planck-Institut für Physik
Föhringer Ring 6, D-80805 München, Germany

**Abstract**

The conservation of the isovector axial current in lattice QCD with massless Wilson quarks is studied to one-loop order of perturbation theory. Following a strategy described in a previous publication, the O($a$) counterterm required for on-shell improvement of the current is computed. We also confirm an earlier result of Wohlert for the coefficient in front of the SW term in the improved action.

June 1996

## 1. Introduction

This paper is the second in a series of publications [1–3] on chiral symmetry and O($a$) improvement in lattice QCD with Wilson quarks. The underlying theoretical framework has been explained in detail in ref. [1]. We now report on a study of the conservation of the isovector axial current to one-loop order of perturbation theory. Our aim is to verify that the O($a$) improvement of the lattice action and the current works out in the expected way. In particular, we shall show that the coefficients of the associated O($a$) counterterms, $c_{\rm sw}$ and $c_{\rm A}$, can be computed by requiring the lattice corrections to the PCAC relation to be of order $a^2$ (as suggested in ref. [1]).

The coefficient $c_{\rm sw}$ of the SW term [4] in the improved action has been calculated to one-loop order many years ago by Wohlert [5]. His calculation has partly been repeated by Naik [7], but it is only now that a complete and independent check of Wohlert's result is obtained. The other coefficient, $c_{\rm A}$, vanishes at tree-level and a one-loop calculation is hence required for a first estimate of the magnitude of the associated O($a$) counterterm.

Throughout this paper we assume that the reader is familiar with ref. [1]. The notations introduced there are taken over completely without further notice. Equations in ref. [1] are referred to by prefixing a Roman "I" to the equation number.

The correlation functions involving the axial current that we shall consider are constructed from the Schrödinger functional. We first need to discuss the evaluation of the quark functional integral in this framework, which is not totally trivial because the boundary conditions are rather special (sect. 2). In sect. 3 the PCAC relation and the approach to the continuum limit are studied in the free quark theory. The perturbation expansion of the relevant correlation functions in the full theory is derived in sect. 4. We are then in a position to compute the one-loop coefficients $c_{\rm A}^{(1)}$ and $c_{\rm sw}^{(1)}$ (sects. 5,6). In the last section we show that the lattice effects that remain after improvement are rather small in general.

We are indebted to our colleagues Andrea Galli, Stefan Sint, Rainer Sommer and Ulli Wolff for their help at various stages of this work. We also thank the computer centers at CERN and EPFL for allowing us to use their SP2 and T3D machines.

## 2. Quark propagation

In this section we consider the $\mathrm{O}(a)$ improved Schrödinger functional defined in sects. 4 and 5 of ref. [1] and work out the integral over the quark and anti-quark fields in some detail.

*2.1 Lattice Dirac operator*

For all times $x_0$ in the range $0 < x_0 < T$ we have

$$\frac{\delta S_{\mathrm{impr}}}{\delta\overline{\psi}(x)} = (D + \delta D + m_0)\psi(x), \tag{2.1}$$

where $D$ denotes the Wilson-Dirac operator (I.2.3). The $\mathrm{O}(a)$ correction $\delta D$ is a sum of a volume and a boundary term,

$$\delta D = \delta D_{\mathrm{v}} + \delta D_{\mathrm{b}}. \tag{2.2}$$

Explicitly the volume term is given by

$$\delta D_{\mathrm{v}}\psi(x) = c_{\mathrm{sw}}\tfrac{i}{4}a\sigma_{\mu\nu}\widehat{F}_{\mu\nu}(x)\psi(x), \tag{2.3}$$

while for the boundary term one obtains

$$\begin{aligned}\delta D_{\mathrm{b}}\psi(x) = (\tilde{c}_{\mathrm{t}} - 1)\frac{1}{a}\Big\{&\delta_{x_0,a}\left[\psi(x) - U(x - a\hat{0}, 0)^{-1}P_{+}\psi(x - a\hat{0})\right] \\ &+ \delta_{x_0,T-a}\left[\psi(x) - U(x, 0)P_{-}\psi(x + a\hat{0})\right]\Big\}.\end{aligned} \tag{2.4}$$

Note that in this expression there is no reference to the "undefined" components of the quark field at the boundaries of the lattice. $D + \delta D$ is hence a well-defined operator acting on quark fields satisfying Schrödinger functional boundary conditions.

*2.2 Quark propagator and classical solutions*

In the presence of a given gauge field $U(x, \mu)$, the quark propagator $S(x, y)$ is defined through

$$(D + \delta D + m_0)S(x, y) = a^{-4}\delta_{xy}, \qquad 0 < x_0 < T, \tag{2.5}$$

and the boundary conditions

$$P_{+}S(x,y)|_{x_0=0} = P_{-}S(x,y)|_{x_0=T} = 0. \tag{2.6}$$

We assume that the gauge field and the quark mass are such that the solution of eq. (2.5) is unique. As in infinite volume one can show that

$$S(x,y)^{\dagger} = \gamma_5 S(y,x)\gamma_5. \tag{2.7}$$

An analytic expression for the propagator in the free quark theory will be given in sect. 3.

The solution of the Dirac equation

$$(D + \delta D + m_0)\psi_{\rm cl}(x) = 0, \qquad 0 < x_0 < T, \tag{2.8}$$

with boundary values

$$P_{+}\psi_{\rm cl}(x)|_{x_0=0} = \rho(\mathbf{x}), \qquad P_{-}\psi_{\rm cl}(x)|_{x_0=T} = \rho'(\mathbf{x}), \tag{2.9}$$

plays an important rôle in the following. For $0 < x_0 < T$ it is given by

$$\psi_{\rm cl}(x) = a^3 \sum_{\mathbf{y}} \tilde{c}_{\rm t}\big\{S(x,y)U(y-a\hat{0},0)^{-1}\rho(\mathbf{y})|_{y_0=a} +$$

$$S(x,y)U(y,0)\rho'(\mathbf{y})|_{y_0=T-a}\big\}. \tag{2.10}$$

It should be emphasized that this representation is not valid if $x_0 = 0$ or $x_0 = T$. We shall also make use of the solution of the adjoint equation

$$\overline{\psi}_{\rm cl}(x)(\overleftarrow{D}^{\dagger} + \delta\overleftarrow{D}^{\dagger} + m_0) = 0, \qquad 0 < x_0 < T, \tag{2.11}$$

with boundary values

$$\overline{\psi}_{\rm cl}(x)P_{-}|_{x_0=0} = \bar{\rho}(\mathbf{x}), \qquad \overline{\psi}_{\rm cl}(x)P_{+}|_{x_0=T} = \bar{\rho}'(\mathbf{x}). \tag{2.12}$$

The adjoint action of the Dirac operator in eq. (2.11) is defined through

$$-\frac{\delta S_{\rm impr}}{\delta\psi(x)} = \overline{\psi}(x)(\overleftarrow{D}^{\dagger} + \delta\overleftarrow{D}^{\dagger} + m_0), \tag{2.13}$$

and it is again possible to express the solution through the propagator in a way analogous to eq. (2.10).

*2.3 Quark functional integral*

We are interested in evaluating the expectation value of polynomials $\mathcal{O}$ in the gauge field variables, the quark and anti-quark fields in the interior of the space-time volume and the boundary fields $\zeta(\mathbf{y}),\ldots,\bar{\zeta}'(\mathbf{z})$. If we first integrate over the quark fields, the expectation value assumes the form

$$\langle\mathcal{O}\rangle = \langle\,[\mathcal{O}]_{\rm F}\,\rangle_{\rm G}, \tag{2.14}$$

where $\langle\ldots\rangle_{\rm G}$ denotes the gauge field average with probability density proportional to

$$\det(D+\delta D+m_0)\exp\{-S_{\rm G}[U]-\delta S_{\rm G,b}[U]\}. \tag{2.15}$$

The Dirac operator is here considered to be a linear mapping in the space of all quark fields with zero boundary values.

In the following we deduce a set of simple rules that allow us to calculate the quark field average $[\mathcal{O}]_{\rm F}$. We begin by introducing the generating functional

$$\mathcal{Z}_{\rm F}[\bar{\rho}\,',\rho\,';\bar{\rho},\rho;\bar{\eta},\eta;U] = \int {\rm D}[\,\psi\,]{\rm D}[\,\overline{\psi}\,]\,\exp\bigl\{-S_{\rm F,impr}[U,\overline{\psi},\psi\,] + a^4\sum_x\,\bigl[\,\overline{\psi}(x)\eta(x)+\bar{\eta}(x)\psi(x)\bigr]\bigr\}. \tag{2.16}$$

The improved quark action appearing in this formula is equal to the sum of all terms in the improved action (I.5.2) that depend on the quark fields. $\eta(x)$ and $\bar{\eta}(x)$, $0<x_0<T$, are source fields for the quark and anti-quark fields in the interior of the space-time volume. If we substitute

$$\psi(x)\rightarrow\frac{\delta}{\delta\bar{\eta}(x)}, \qquad \overline{\psi}(x)\rightarrow-\frac{\delta}{\delta\eta(x)}, \tag{2.17}$$

in the polynomial $\mathcal{O}$, its quark field average is obtained through

$$[\mathcal{O}]_{\rm F} = \left\{\frac{1}{\mathcal{Z}_{\rm F}}\mathcal{O}\mathcal{Z}_{\rm F}\right\}_{\bar{\rho}\,'=\ldots=\eta=0}. \tag{2.18}$$

As will be shown below the generating functional is an exponential of a quadratic expression in the boundary values $\bar{\rho}\,',\ldots,\rho$ and the source fields $\bar{\eta},\eta$. The differentiations in eq. (2.18) may hence be worked out by applying Wick's theorem, i.e. $[\mathcal{O}]_{\rm F}$ is expressed as a sum of Wick contractions. We are then left with the task to list all quark two-point functions.

To this end we insert the decomposition

$$\psi(x) = \psi_{\rm cl}(x) + \chi(x), \qquad \overline{\psi}(x) = \overline{\psi}_{\rm cl}(x) + \overline{\chi}(x), \tag{2.19}$$

in the functional integral (2.16). The quantum components of the quark fields, $\overline{\chi}(x)$ and $\chi(x)$, have vanishing boundary values. It is then easy to show that

$$S_{\rm F,impr}[U,\overline{\psi},\psi] = S_{\rm F,impr}[U,\overline{\psi}_{\rm cl},\psi_{\rm cl}] + S_{\rm F,impr}[U,\overline{\chi},\chi], \tag{2.20}$$

and for the generating functional we thus obtain

$$\begin{aligned} \ln \mathcal{Z}_{\rm F} &= \ln \mathcal{Z}_{\rm F}|_{\bar{\rho}'=\ldots=\eta=0} - S_{\rm F,impr}[U,\overline{\psi}_{\rm cl},\psi_{\rm cl}] \\ &\quad + a^8 \sum_{x,y} \bar{\eta}(x) S(x,y) \eta(y) + a^4 \sum_{x} \left[\bar{\eta}(x)\psi_{\rm cl}(x) + \overline{\psi}_{\rm cl}(x)\eta(x)\right]. \end{aligned} \tag{2.21}$$

The second term on the right hand side of this equation may be simplified by noting that $(D + \delta D + m_0)\psi_{\rm cl}(x)$ vanishes for $0 < x_0 < T$. Taking the boundary conditions into account, some algebra then yields

$$\begin{aligned} &S_{\rm F,impr}[U,\overline{\psi}_{\rm cl},\psi_{\rm cl}] = \\ &a^3 \sum_{\mathbf{x}} \Big\{ \tfrac{1}{2} a \tilde{c}_{\rm s} \Big[\bar{\rho}(\mathbf{x})\gamma_k(\nabla_k^* + \nabla_k)\rho(\mathbf{x}) + \bar{\rho}'(\mathbf{x})\gamma_k(\nabla_k^* + \nabla_k)\rho'(\mathbf{x})\Big] \\ &- \tilde{c}_{\rm t} \Big[\bar{\rho}(\mathbf{x}) U(x - a\hat{0}, 0)\psi_{\rm cl}(x)\big|_{x_0=a} + \bar{\rho}'(\mathbf{x}) U(x,0)^{-1}\psi_{\rm cl}(x)\big|_{x_0=T-a}\Big]\Big\}. \end{aligned} \tag{2.22}$$

The basic two-point functions may now be calculated by differentiating with respect to the boundary values and the source fields. The complete list of non-zero contractions is

$$\left[\psi(x)\overline{\psi}(y)\right]_{\rm F} = S(x,y), \tag{2.23}$$

$$\left[\psi(x)\bar{\zeta}(\mathbf{y})\right]_{\rm F} = \tilde{c}_{\rm t}\; S(x,y)U(y - a\hat{0}, 0)^{-1}P_+\big|_{y_0=a}, \tag{2.24}$$

$$\left[\psi(x)\bar{\zeta}'(\mathbf{y})\right]_{\rm F} = \tilde{c}_{\rm t}\; S(x,y)U(y,0)P_-\big|_{y_0=T-a}, \tag{2.25}$$

$$\left[\zeta(\mathbf{x})\overline{\psi}(y)\right]_{\rm F} = \tilde{c}_{\rm t}\; P_{-}U(x-a\hat{0},0)S(x,y)\big|_{x_0=a}\,, \tag{2.26}$$

$$\left[\zeta\,'(\mathbf{x})\overline{\psi}(y)\right]_{\rm F} = \tilde{c}_{\rm t}\; P_{+}U(x,0)^{-1}S(x,y)\big|_{x_0=T-a}\,, \tag{2.27}$$

$$\begin{aligned}\left[\zeta(\mathbf{x})\bar{\zeta}(\mathbf{y})\right]_{\rm F} &= \tilde{c}_{\rm t}^2\; P_{-}U(x-a\hat{0},0)S(x,y)U(y-a\hat{0},0)^{-1}P_{+}\big|_{x_0=y_0=a} \\ &\quad - \tfrac{1}{2}\tilde{c}_{\rm s}P_{-}\gamma_k(\nabla_k^{*}+\nabla_k)a^{-2}\delta_{\mathbf{xy}},\end{aligned} \tag{2.28}$$

$$\left[\zeta(\mathbf{x})\bar{\zeta}\,'(\mathbf{y})\right]_{\rm F} = \tilde{c}_{\rm t}^2\; P_{-}U(x-a\hat{0},0)S(x,y)U(y,0)^{-1}P_{-}\big|_{x_0=a,y_0=T-a}\,, \tag{2.29}$$

$$\left[\zeta\,'(\mathbf{x})\bar{\zeta}(\mathbf{y})\right]_{\rm F} = \tilde{c}_{\rm t}^2\; P_{+}U(x,0)^{-1}S(x,y)U(y-a\hat{0},0)^{-1}P_{+}\big|_{x_0=T-a,y_0=a}\,, \tag{2.30}$$

$$\begin{aligned}\left[\zeta\,'(\mathbf{x})\bar{\zeta}\,'(\mathbf{y})\right]_{\rm F} &= \tilde{c}_{\rm t}^2\; P_{+}U(x,0)^{-1}S(x,y)U(y,0)P_{-}\big|_{x_0=y_0=T-a} \\ &\quad - \tfrac{1}{2}\tilde{c}_{\rm s}P_{+}\gamma_k(\nabla_k^{*}+\nabla_k)a^{-2}\delta_{\mathbf{xy}}.\end{aligned} \tag{2.31}$$

We have here made use of eq. (2.10) and the analogous representation of $\overline{\psi}_{\rm cl}(x)$ to express all contractions through the quark propagator.

*2.4 Example*

To illustrate the procedure we consider the correlation functions $f_{\rm A}(x_0)$ and $f_{\rm P}(x_0)$ introduced in subsect. 6.3 of ref. [1]. Applying Wick's theorem one deduces that

$$f_{\rm A}(x_0) = a^6\sum_{\mathbf{y},\mathbf{z}}\tfrac{1}{2}\left\langle {\rm tr}\left\{\left[\zeta(\mathbf{z})\overline{\psi}(x)\right]_{\rm F}\gamma_0\gamma_5\left[\psi(x)\bar{\zeta}(\mathbf{y})\right]_{\rm F}\gamma_5\right\}\right\rangle_{\rm G}, \tag{2.32}$$

$$f_{\rm P}(x_0) = a^6\sum_{\mathbf{y},\mathbf{z}}\tfrac{1}{2}\left\langle {\rm tr}\left\{\left[\zeta(\mathbf{z})\overline{\psi}(x)\right]_{\rm F}\gamma_5\left[\psi(x)\bar{\zeta}(\mathbf{y})\right]_{\rm F}\gamma_5\right\}\right\rangle_{\rm G}, \tag{2.33}$$

where the contractions are given by eqs. (2.24),(2.26) and the trace is to be taken over the Dirac and colour indices only. Note that, as a consequence of the hermiticity property (2.7), we have

$$\gamma_5\left[\zeta(\mathbf{z})\overline{\psi}(x)\right]_{\rm F}\gamma_5 = \left\{\left[\psi(x)\bar{\zeta}(\mathbf{z})\right]_{\rm F}\right\}^{\dagger}. \tag{2.34}$$

The correlation function $f_{\rm P}(x_0)$ is hence an average of a manifestly positive quantity.

## 3. Free quark theory

We now proceed to study the propagation of free Wilson quarks with Schrödinger functional boundary conditions. The formalism developed above applies in this case too with the obvious modifications. We set $\tilde{c}_{\rm s} = \tilde{c}_{\rm t} = 1$ in this section and shall justify this choice in subsect. 3.3.

*3.1 Analytic expression for the quark propagator*

To introduce our notations we first discuss the plane wave solutions of Dirac's equation on lattices that are infinitely extended in the time direction. It suffices to consider the positive energy solutions

$$\psi(x) = u\,{\rm e}^{ipx}, \quad {\rm Im}\, p_0 \geq 0. \tag{3.1}$$

The spatial components of the momentum $p$ are in the range

$$-\pi/a < p_k \leq \pi/a. \tag{3.2}$$

They must be integer multiples of $2\pi/L$ to satisfy the boundary conditions in the space directions. The constant spinor $u$ and the energy $p_0$ are constrained by Dirac's equation, $(D + m_0)\psi(x) = 0$, which reduces to

$$\{i\gamma_\mu \mathring{p}_\mu^+ + M(p^+)\}\, u = 0. \tag{3.3}$$

Here and below we use the abbreviations

$$\mathring{q}_\mu = (1/a)\sin(aq_\mu)\,, \tag{3.4}$$

$$\hat{q}_\mu = (2/a)\sin(aq_\mu/2)\,, \tag{3.5}$$

for any momentum $q$. The effective mass in eq. (3.3) is given by

$$M(q) = m_0 + \tfrac{1}{2}a\hat{q}^2, \tag{3.6}$$

and the momentum $p_\mu^+ = p_\mu + \theta_\mu/L$ includes a shift by the angles $\theta_\mu$ (cf. subsect. 4.2 and appendix A of ref. [1]).

Eq. (3.3) is similar to Dirac's equation in the continuum theory and its solution is obtained in essentially the same manner. We first multiply the equation from the left with $i\gamma_\mu \mathring{p}^+_\mu - M(p^+)$ and deduce that

$$(\mathring{p}^+)^2 + M(p^+)^2 = 0. \tag{3.7}$$

Solving for the energy $p_0$ one obtains

$$p_0 = p_0^+ = i\omega(\mathbf{p}^+) \bmod 2\pi/a, \tag{3.8}$$

where $\omega(\mathbf{q})$ is given by

$$\sinh\left[\frac{a}{2}\omega(\mathbf{q})\right] = \frac{a}{2}\left\{\frac{\mathring{\mathbf{q}}^2 + (m_0 + \frac{1}{2}a\hat{\mathbf{q}}^2)^2}{1 + a(m_0 + \frac{1}{2}a\hat{\mathbf{q}}^2)}\right\}^{\frac{1}{2}}. \tag{3.9}$$

$\omega(\mathbf{p}^+)$ is well-defined and non-negative for $m_0 \geq 0$. There are no other positive energy solutions of eq. (3.7).

It remains to determine the spinor $u$. With $p_0$ as above, eq. (3.3) has two linearly independent solutions, corresponding to spin up and spin down states. Explicit expressions could be given as in the continuum theory, but we shall not need them here.

The propagator $S(x, y)$ can now be constructed following standard procedures. We omit the details and state the result in the form

$$S(x, y) = (D^\dagger + m_0)G(x, y), \qquad 0 < x_0, y_0 < T, \tag{3.10}$$

where $G(x, y)$ is defined through

$$\begin{aligned} G(x, y) = L^{-3} \sum_{\mathbf{p}} &\{-2i\mathring{p}_0^+ A(\mathbf{p}^+)R(p^+)\}^{-1}\, \mathrm{e}^{i\mathbf{p}(\mathbf{x}-\mathbf{y})} \\ \times \Big\{ &(M(p^+) - i\mathring{p}_0^+)\mathrm{e}^{-\omega(\mathbf{p}^+)|x_0-y_0|} + (M(p^+) + i\mathring{p}_0^+)\mathrm{e}^{-\omega(\mathbf{p}^+)(2T-|x_0-y_0|)} \\ - &(M(p^+) + i\gamma_0\mathring{p}_0^+)\mathrm{e}^{-\omega(\mathbf{p}^+)(x_0+y_0)} - (M(p^+) - i\gamma_0\mathring{p}_0^+)\mathrm{e}^{-\omega(\mathbf{p}^+)(2T-x_0-y_0)}\Big\}, \end{aligned} \tag{3.11}$$

for all times $x_0$ (including 0 and $T$). The amplitudes $A$ and $R$ occurring in this expression are given by

$$A(\mathbf{q}) = 1 + a(m_0 + \tfrac{1}{2}a\hat{\mathbf{q}}^2), \tag{3.12}$$

$$R(q) = M(q)\left\{1 - \mathrm{e}^{-2\omega(\mathbf{q})T}\right\} - i\mathring{q}_0\left\{1 + \mathrm{e}^{-2\omega(\mathbf{q})T}\right\}. \qquad (3.13)$$

The sum runs over all momenta $\mathbf{p}$ in the range (3.2) and $p^+$ is on the mass-shell (3.8). Note that $R(p^+)$ vanishes if $\mathbf{p}^+ = \mathbf{0}$ and $m_0 = 0$. The propagator remains well-defined, however, and can be calculated by first assuming $m_0$ to be positive and then taking the limit $m_0 \to 0$.

*3.2 Axial current conservation*

As explained in sect. 6 of ref. [1] the PCAC relation in the improved theory is violated by cutoff effects of order $a^2$. Without improvement the effects are of order $a$ in general. The PCAC relation may hence be used as a test for improvement. Specifically it has been proposed to introduce an unrenormalized current quark mass $m$ through eq. (I.6.13). Any variation of $m$ as a function of the kinematical parameters $T$, $L$, etc., can be taken as a measure for the size of the lattice effects at the chosen lattice spacing.

In the free quark theory the $\mathrm{O}(a)$ counterterm $\delta A_\mu^a(x)$ to the axial current is expected to vanish [6] and $m$ is then given by

$$m = \tfrac{1}{4}(\partial_0^* + \partial_0) f_{\mathrm{A}}(x_0)/f_{\mathrm{P}}(x_0). \qquad (3.14)$$

We are now in a position to compute this ratio analytically. Combining eqs. (2.32),(2.33) with (2.24),(2.26) and the exact expression for the free propagator, one obtains (after some algebra)

$$f_{\mathrm{A}}(x_0) = \frac{2N}{R(p^+)^2}\Big\{2M_+(p^+)M_-(p^+)\mathrm{e}^{-2\omega(\mathbf{p}^+)T} - M(p^+)\Big[M_-(p^+)\mathrm{e}^{-2\omega(\mathbf{p}^+)x_0} + M_+(p^+)\mathrm{e}^{-2\omega(\mathbf{p}^+)(2T-x_0)}\Big]\Big\}, \qquad (3.15)$$

$$f_{\mathrm{P}}(x_0) = -\frac{2i\mathring{p}_0^+ N}{R(p^+)^2}\Big\{M_-(p^+)\mathrm{e}^{-2\omega(\mathbf{p}^+)x_0} - M_+(p^+)\mathrm{e}^{-2\omega(\mathbf{p}^+)(2T-x_0)}\Big\}, \qquad (3.16)$$

where $N$ denotes the number of colours, $\mathbf{p} = \mathbf{0}$ and

$$M_\pm(p^+) = M(p^+) \pm i\mathring{p}_0^+ \qquad (3.17)$$

[as before the momentum $p^+$ is assumed to be on the mass-shell (3.8)]. From these formulae one immediately deduces that

$$m = M(p^+)\cosh[a\omega(\mathbf{p}^+)] \tag{3.18}$$

for all $x_0$ in the range $a < x_0 < T - a$.

It may come as a surprise that $m$ turns out to be independent of $x_0$. Although this is the expected behaviour in the continuum limit, lattice effects in general imply that $m$ is a non-trivial function of $x_0$. If there were no lattice effects at all, $m$ would also have to be independent of the kinematical parameters such as $L$ and the angles $\theta_k$. This is evidently not the case, but from the above it follows that

$$m = m_{\mathrm{p}} + \mathrm{O}(a^2), \tag{3.19}$$

where $m_{\mathrm{p}}$ denotes the pole mass (I.3.6). In other words, there is no dependence on the kinematical parameters at order $a$.

### *3.3 Continuum limit and O(a) improvement*

Since all correlation functions are obtained as products of the basic two-point functions (2.23)–(2.31), it suffices to study the approach to the continuum limit of the latter. Before taking $a \to 0$ we must specify in which way the quark mass is to be treated. Our discussion above suggests that we should keep the pole mass $m_{\mathrm{p}}$ fixed instead of the bare mass $m_0$. The name "pole mass" derives from the observation that

$$\omega(\mathbf{0}) = |m_{\mathrm{p}}|. \tag{3.20}$$

For general momenta $\mathbf{q}$ we have

$$\omega(\mathbf{q}) = \epsilon(\mathbf{q}) + \mathrm{O}(a^2), \qquad \epsilon(\mathbf{q}) = \sqrt{m_{\mathrm{p}}^2 + \mathbf{q}^2}. \tag{3.21}$$

It is also straightforward to show that

$$M(p^+) = m_{\mathrm{p}} + \mathrm{O}(a^2), \tag{3.22}$$

$$R(p^+) = m_{\mathrm{p}}\left\{1 - \mathrm{e}^{-2\epsilon(\mathbf{p}^+)T}\right\} + \epsilon(\mathbf{p}^+)\left\{1 + \mathrm{e}^{-2\epsilon(\mathbf{p}^+)T}\right\} + \mathrm{O}(a^2), \tag{3.23}$$

for any momentum $p^+$ on the mass-shell (3.8). Using the explicit expression for the quark propagator, the continuum limit of the contractions (2.23)–(2.31) is now easily worked out.

We first consider the two-point functions of the boundary fields,

$$a^3 \sum_{\mathbf{x}} \mathrm{e}^{-i\mathbf{p}(\mathbf{x}-\mathbf{y})} \left\langle \zeta(\mathbf{x})\bar{\zeta}(\mathbf{y}) \right\rangle = -i\gamma_k \mathring{p}_k^+ \frac{A(\mathbf{p}^+)}{R(p^+)} \left\{ 1 - \mathrm{e}^{-2\omega(\mathbf{p}^+)T} \right\} P_+, \quad (3.24)$$

$$a^3 \sum_{\mathbf{x}} \mathrm{e}^{-i\mathbf{p}(\mathbf{x}-\mathbf{y})} \left\langle \zeta(\mathbf{x})\bar{\zeta}'(\mathbf{y}) \right\rangle = -2i\mathring{p}_0^+ \frac{A(\mathbf{p}^+)}{R(p^+)} \mathrm{e}^{-\omega(\mathbf{p}^+)T} P_-. \quad (3.25)$$

Since $A(\mathbf{p}^+) = 1 + am_0 + \mathrm{O}(a^2)$, it follows that these correlation functions are $\mathrm{O}(a)$ improved up to an overall normalization factor equal to $1 + am_0$. Note that the second term in eq. (2.28) cancels a momentum dependent contribution of order $a$. Without this term the correlation function (3.24) would not be improved.

The mixed two-point functions, involving a boundary field and the bulk field $\psi(x)$ or $\overline{\psi}(x)$, are also $\mathrm{O}(a)$ improved. This is immediately clear from

$$a^3 \sum_{\mathbf{x}} \mathrm{e}^{-i\mathbf{p}(\mathbf{x}-\mathbf{y})} \left\langle \psi(x)\bar{\zeta}(\mathbf{y}) \right\rangle = \frac{1}{R(p^+)} \Big\{ (M(p^+) - i\mathring{p}_0^+ - i\gamma_k \mathring{p}_k^+) \mathrm{e}^{-\omega(\mathbf{p}^+)x_0}$$

$$- (M(p^+) + i\mathring{p}_0^+ - i\gamma_k \mathring{p}_k^+) \mathrm{e}^{-\omega(\mathbf{p}^+)(2T - x_0)} \Big\} P_+. \quad (3.26)$$

No renormalization factor of the form $1 + \mathrm{O}(am_0)$ is found here, because the factors associated with the boundary and bulk fields are inverse to each other and thus cancel in the mixed two-point functions. An $\mathrm{O}(a)$ normalization factor equal to $(1 + am_0)^{-1}$ in fact shows up in the correlation function $\langle \psi(x)\overline{\psi}(y) \rangle$, as one may easily verify from eqs. (3.10),(3.11).

We thus conclude that the continuum limit of the free quark Schrödinger functional is reached with a rate proportional to $a^2$, provided the fields are renormalized by factors of the form $1 + b(g_0^2)am_{\mathrm{q}}$ and if we keep the pole mass $m_{\mathrm{p}}$ fixed instead of the bare mass (cf. sects. 2,3 of ref. [1]). In particular, the PCAC relation holds up to corrections of order $a^2$. If we expand the $b$–coefficients according to eq. (I.3.17), our calculations imply that

$$b_{\mathrm{m}}^{(0)} = b_{\zeta}^{(0)} = -\tfrac{1}{2}. \quad (3.27)$$

We can also easily take the continuum limit of $f_{\mathrm{A}}(x_0)$ and $f_{\mathrm{P}}(x_0)$ and verify that these correlation functions are improved if we set

$$b_{\mathrm{A}}^{(0)} = b_{\mathrm{P}}^{(0)} = 1, \quad (3.28)$$

thus confirming a result originally obtained in ref. [6].

As anticipated the $O(a)$ boundary counterterms in the improved quark action are not needed for $O(a)$ improvement at tree-level. One may in fact prove that they must be set to zero as otherwise one would end up with uncancelled $O(a)$ terms in some of the correlation functions considered above. The same is true for the $O(a)$ correction to the axial current, i.e. the coefficient $c_{\rm A}$ must vanish at $g_0 = 0$.

## 4. Expansion of $f_{\rm A}(x_0)$ and $f_{\rm P}(x_0)$ to one-loop order

To determine the coefficients $c_{\rm A}$ and $c_{\rm sw}$ to one-loop order of perturbation theory we need to compute the correlation functions $f_{\rm A}(x_0)$ and $f_{\rm P}(x_0)$ to order $g_0^2$ for various choices of the kinematical parameters. In this section we describe in some detail how this calculation is done for the case of vanishing boundary values $C$ and $C'$ of the gauge field. The modifications required for non-zero boundary values will be discussed in sect. 6.

With little additional effort we shall also be able to compute the one-loop coefficient $\tilde{c}_{\rm t}^{(1)}$. The associated $O(a)$ boundary counterterm is hence not dropped here, although we shall not need to include it when calculating the other coefficients (cf. eq. (I.5.25) and subsect. 6.1 of ref. [1]).

### *4.1 Preliminaries*

To evaluate the expectation values (2.32) and (2.33) in perturbation theory, we first fix the gauge and then expand the action and the quark field contractions in powers of the coupling. As for the action we shall not need to go beyond the leading term. The expansion of the two-point contractions can be simplified by noting that

$$\left[\psi(x)\bar{\zeta}(\mathbf{y})\right]_{\rm F} = \frac{\delta\psi_{\rm cl}(x)}{\delta\rho(\mathbf{y})}. \tag{4.1}$$

So if we define the matrix

$$H(x) = a^3 \sum_{\mathbf{y}} \frac{\delta\psi_{\rm cl}(x)}{\delta\rho(\mathbf{y})}, \tag{4.2}$$

and recall eq. (2.34), it follows that

$$f_{\rm X}(x_0) = \tfrac{1}{2}\left\langle \operatorname{tr}\left\{H(x)^\dagger \Gamma_{\rm X} H(x)\right\}\right\rangle_{\rm G} \tag{4.3}$$

with $\Gamma_{\rm A} = -\gamma_0$ and $\Gamma_{\rm P} = 1$. It is, therefore, sufficient to work out the perturbation expansion of $\psi_{\rm cl}(x)$ to order $g_0^2$.

*4.2 Gauge fixing and the gluon propagator*

The calculation of the expectation value in eq. (4.3) involves an integration over all lattice gauge fields with the specified boundary values. Under a gauge transformation

$$U(x,\mu) \to \Omega(x)U(x,\mu)\Omega(x+a\hat{\mu})^{-1}, \qquad \Omega(x) \in {\rm SU}(N), \tag{4.4}$$

the integrand and the boundary values are invariant, provided the gauge function $\Omega(x)$ is independent of $\mathbf{x}$ at the boundaries $x_0 = 0$ and $x_0 = T$.

To fix this gauge degeneracy we may follow the steps taken in sect. 6 of ref. [9] for the analogous case of the Schrödinger functional in pure gauge theories. A subtle point to note is that the boundary conditions on the ghost fields $c(x)$ and $\bar{c}(x)$ are not the same for all boundary values of the gauge field. For $C = C' = 0$ the boundary conditions are that the ghost fields should vanish at $x_0 = T$, while at $x_0 = 0$ they may be non-zero but should be independent of $\mathbf{x}$. The linear space of all fields with these boundary conditions [and taking values in the Lie algebra of SU($N$)] is denoted by $\mathcal{L}$.

In the gauge fixed theory the link variables are parametrized through a vector field $q_\mu(x)$ according to

$$U(x,\mu) = \exp\{g_0 a q_\mu(x)\}. \tag{4.5}$$

Lattice vector fields $q_\mu(x)$ are defined for $0 \le x_0 < T$ (if $\mu = 0$) and $0 \le x_0 \le T$ (otherwise). They take values in the Lie algebra of SU($N$) and satisfy the boundary conditions

$$q_k(x)|_{x_0=0} = q_k(x)|_{x_0=T} = 0. \tag{4.6}$$

The space of all these fields is denoted by $\mathcal{H}$.

To be able to write the gauge fixing term in a compact form we introduce the operator

$$d:\ \mathcal{L} \mapsto \mathcal{H}, \qquad (d\omega)_\mu(x) = \partial_\mu \omega(x). \tag{4.7}$$

The adjoint $d^*$ of $d$ maps any vector field $q \in \mathcal{H}$ onto an element of $\mathcal{L}$ such that

$$(d^* q, \omega) = -(q, d\omega) \quad \text{for all} \quad \omega \in \mathcal{L}. \tag{4.8}$$

With the obvious choice for the scalar products on $\mathcal{H}$ and $\mathcal{L}$ we have

$$d^*q(x) = \begin{cases} \partial_\mu^* q_\mu(x) & \text{if } 0 < x_0 < T, \\ (a^2/L^3)\sum_{\mathbf{y}} q_0(y)|_{y_0=0} & \text{if } x_0 = 0, \\ 0 & \text{if } x_0 = T. \end{cases} \tag{4.9}$$

Note that $d^*q(x)$ satisfies the boundary conditions required for fields in $\mathcal{L}$.

The total action of the gauge fixed theory, including the gauge fixing term and the ghost field action, is now given by

$$\begin{aligned} S_{\rm tot}[q,\bar{c},c] = S_{\rm G}[U] + \delta S_{\rm G,b}[U] - \ {\rm tr}\,\{\ln(D + \delta D + m_0)\} \\ + \tfrac{1}{2}\lambda_0(d^*q, d^*q) - (\bar{c}, d^*\delta_c q), \end{aligned} \tag{4.10}$$

where $\lambda_0$ denotes the bare gauge parameter and $\delta_c q$ the first-order variation of $q$ under the gauge transformation generated by the ghost field $c$. We shall not need the explicit form for $\delta_c q$ here, but note for completeness that

$$\delta_c q_\mu = \partial_\mu c + g_0\,\left[q_\mu, c + \tfrac{1}{2}a\partial_\mu c\right] + {\rm O}(g_0^2) \tag{4.11}$$

(no sum over $\mu$ is implied in this formula). In the following expectation values in the gauge fixed theory, with action (4.10) and a priori measure $\mathrm{D}[U]\mathrm{D}[c]\mathrm{D}[\bar{c}]$, are denoted by $\langle\ldots\rangle_{\tilde{\rm G}}$.

To leading order in the gauge coupling we have

$$S_{\rm tot}[q,\bar{c},c] = \text{constant} + \tfrac{1}{2}(q, \Delta_1 q) + (\bar{c}, \Delta_0 c) + {\rm O}(g_0), \tag{4.12}$$

where $\Delta_0 = -d^*d$ and

$$\Delta_1 = \Delta_1' - \lambda_0 dd^*, \tag{4.13}$$

$$\Delta_1' q_\mu(x) = \sum_{\nu\neq\mu}\{-\partial_\nu^*\partial_\nu q_\mu(x) + \partial_\nu^*\partial_\mu q_\nu(x)\}\,. \tag{4.14}$$

The action of $\Delta_1$ simplifies considerably in the Feynman gauge $\lambda_0 = 1$. Explicitly, for the spatial components one finds

$$\Delta_1 q_k(x) = -\partial_\mu^*\partial_\mu q_k(x), \qquad 0 < x_0 < T, \tag{4.15}$$

while for the time component we have

$$\Delta_1 q_0(x) = \begin{cases} -\partial_\mu^* \partial_\mu q_0(x) & \text{if } 0 < x_0 < T-a, \\ -\partial_k^* \partial_k q_0(x) + a^{-1}\partial_0^* q_0(x) & \text{if } x_0 = T-a, \\ -\partial_k^* \partial_k q_0(x) - a^{-1}\partial_0 q_0(x) & \\ \quad +(a/L^3)\sum_{\mathbf{y}} q_0(y)|_{y_0=0} & \text{if } x_0 = 0. \end{cases} \tag{4.16}$$

It is then straightforward to calculate the gluon propagator

$$\langle q_\mu^a(x) q_\nu^b(y)\rangle_{\tilde{\mathrm{G}}} = \delta^{ab} D_{\mu\nu}(x,y) + \mathrm{O}(g_0^2). \tag{4.17}$$

The result is given in appendix A.

*4.3 Expansion of $\psi_{\rm cl}(x)$ in powers of $g_0$*

Through eq. (4.5) the Wilson-Dirac operator becomes a function of the bare coupling $g_0$. The lowest order term in the expansion

$$D = \sum_{k=0}^{\infty} g_0^k D^{(k)} \tag{4.18}$$

coincides with the free quark Dirac operator studied in sect. 3. For $k \geq 1$ we have

$$D^{(k)}\psi(x) = -\frac{a^{k-1}}{k!}\sum_{\mu=0}^{3}\Big\{\tfrac{1}{2}(1-\gamma_\mu)\lambda_\mu[q_\mu(x)]^k\psi(x+a\hat{\mu})$$
$$+\tfrac{1}{2}(1+\gamma_\mu)\lambda_\mu^{-1}[-q_\mu(x-a\hat{\mu})]^k\psi(x-a\hat{\mu})\Big\}. \tag{4.19}$$

The $\mathrm{O}(a)$ counterterm $\delta D$ can be expanded similarly. In this case the dependence on $g_0$ is also through the coefficients $c_{\rm sw}$ and $\tilde{c}_{\rm t}$. Note that $\delta D$ is of order $g_0$, because the gluon field tensor vanishes proportionally to the coupling and because $\tilde{c}_{\rm t} - 1$ is of order $g_0^2$ (cf. sect. 3).

The classical solution $\psi_{\rm cl}(x)$ may now be obtained in the form of a series

$$\psi_{\rm cl}(x) = \sum_{k=0}^{\infty} g_0^k \psi_{\rm cl}^{(k)}(x), \tag{4.20}$$

where the leading term, $\psi_{\rm cl}^{(0)}(x)$, satisfies the free Wilson-Dirac equation with boundary values $\rho(\mathbf{x})$ and $\rho'(\mathbf{x})$ (cf. subsect. 2.2). For $0 < x_0 < T$ the solution is given by

$$\psi_{\rm cl}^{(0)}(x) = a^3 \sum_{\mathbf{y}} \left\{ S^{(0)}(x,y)\rho(\mathbf{y})|_{y_0=a} + S^{(0)}(x,y)\rho'(\mathbf{y})|_{y_0=T-a} \right\}, \qquad (4.21)$$

with $S^{(0)}(x,y)$ being the free quark propagator discussed in sect. 3. The higher order terms have zero boundary values and may be recursively constructed by solving

$$(D^{(0)} + m_0)\psi_{\rm cl}^{(k)}(x) = -\sum_{j=0}^{k-1} (D^{(k-j)} + \delta D^{(k-j)})\psi_{\rm cl}^{(j)}(x) \qquad (4.22)$$

for $0 < x_0 < T$. In particular, to first and second order we have

$$\psi_{\rm cl}^{(1)}(x) = -a^4 \sum_{y} S^{(0)}(x,y)\tilde{D}^{(1)}\psi_{\rm cl}^{(0)}(y), \qquad (4.23)$$

$$\psi_{\rm cl}^{(2)}(x) = -a^4 \sum_{y} S^{(0)}(x,y)\tilde{D}^{(2)}\psi_{\rm cl}^{(0)}(y) + a^8 \sum_{y,z} S^{(0)}(x,y)\tilde{D}^{(1)}S^{(0)}(y,z)\tilde{D}^{(1)}\psi_{\rm cl}^{(0)}(z), \qquad (4.24)$$

where the abbreviation $\tilde{D}^{(k)} = D^{(k)} + \delta D^{(k)}$ has been used.

### *4.4 Expansion of $f_{\rm X}(x_0)$ and Feynman diagrams*

The perturbation expansion of the matrix $H(x)$ defined in subsect. 4.1 may be obtained straightforwardly by differentiating the series (4.20) with respect to the boundary value $\rho(\mathbf{y})$. To lowest order we have

$$H^{(0)}(x) = a^3 \sum_{\mathbf{y}} \frac{\delta\psi_{\rm cl}^{(0)}(x)}{\delta\rho(\mathbf{y})}, \qquad (4.25)$$

and the higher order terms, $H^{(1)}(x)$ and $H^{(2)}(x)$, are given by eqs. (4.23) and (4.24) with $\psi_{\rm cl}^{(0)}(x)$ replaced by $H^{(0)}(x)$.

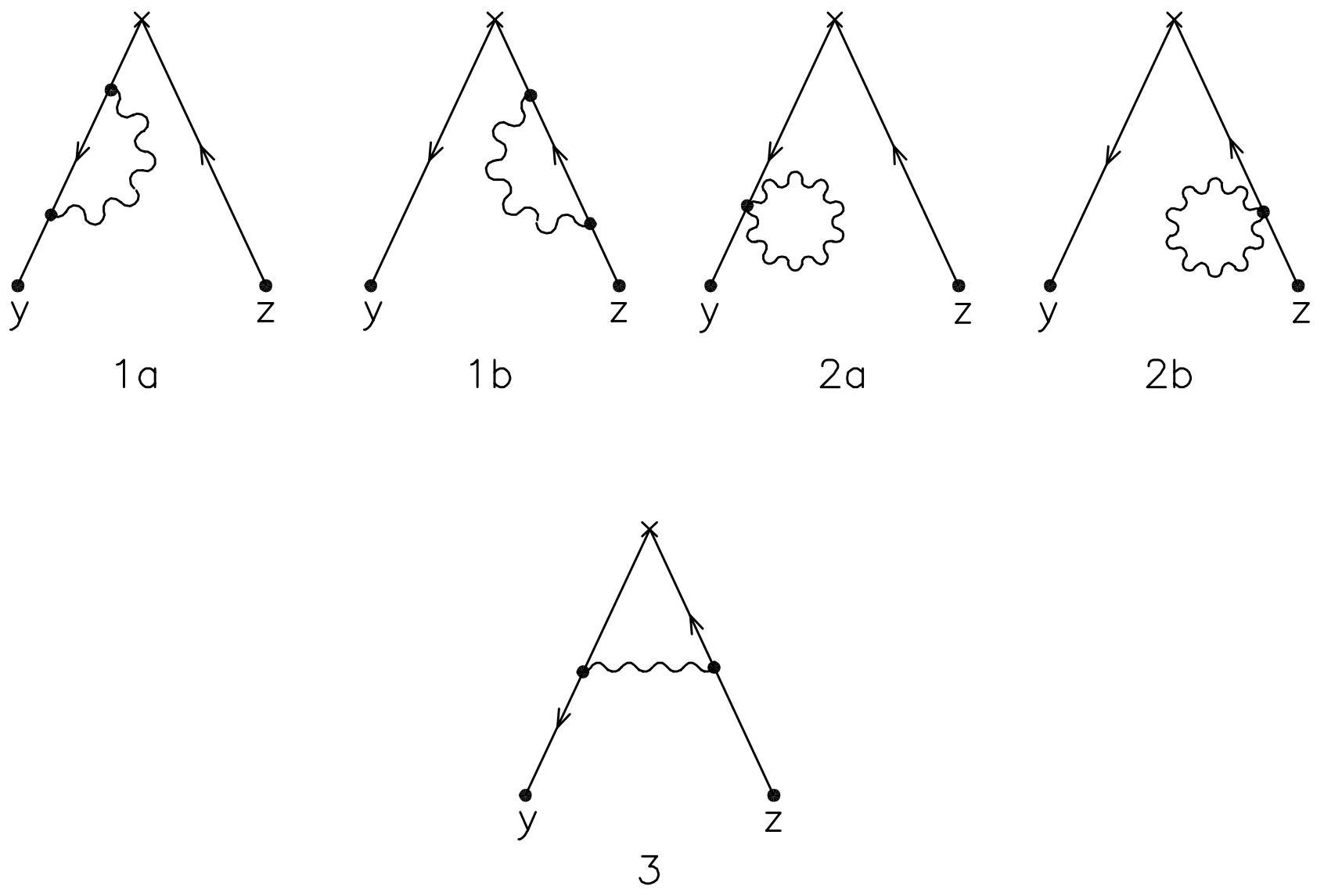

Fig. 1. Feynman diagrams contributing to $f_{\rm X}(x_0)$ at one-loop order of perturbation theory. The cross indicates the insertion of the axial current or density, while the terminal points of the quark lines are at the boundary $y_0=z_0=0$ of the lattice.

If we now insert the expansion of $H(x)$ in eq. (4.3) and use $\langle 1\rangle_{\tilde{\rm G}}=1$ and $\langle q_\mu(x)\rangle_{\tilde{\rm G}}=0$, it follows that

$$f_{\rm X}(x_0)=\sum_{k=0}^{\infty} g_0^{2k} f_{\rm X}^{(k)}(x_0), \tag{4.26}$$

$$f_{\rm X}^{(0)}(x_0)=\tfrac{1}{2}\,{\rm tr}\big\{H^{(0)}(x)^\dagger\Gamma_{\rm X}H^{(0)}(x)\big\}, \tag{4.27}$$

$$\begin{aligned} f_{\rm X}^{(1)}(x_0)&=\tfrac{1}{2}\big\langle\,{\rm tr}\big\{H^{(1)}(x)^\dagger\Gamma_{\rm X}H^{(1)}(x)\big\}\big\rangle_{\tilde{\rm G}}\\ &+\tfrac{1}{2}\big\langle\,{\rm tr}\big\{H^{(2)}(x)^\dagger\Gamma_{\rm X}H^{(0)}(x)+H^{(0)}(x)^\dagger\Gamma_{\rm X}H^{(2)}(x)\big\}\big\rangle_{\tilde{\rm G}}. \end{aligned} \tag{4.28}$$

The expectation values in the last equation are to be taken at $g_0=0$. Note that $H^{(1)}(x)$ is proportional to the gluon field $q_\mu(y)$. After contracting the gluon fields, the first term on the right hand side of eq. (4.28) may thus be represented by the diagram no. 3 in fig. 1. As usual the free quark propagators are depicted by directed solid lines, while the wiggly line stands for the gluon propagator. The quark gluon vertex may be inferred from the operator $\tilde{D}^{(1)}$.

The second order correction $H^{(2)}(x)$ is a sum of several contributions. Most of them are proportional to $q^a_\mu(y)q^b_\nu(z)$ and after contracting these fields one ends up with the diagrams no. 1 and 2 in fig. 1. There is one more term,

$$H^{(2)}(x)_{\rm b} = -a^4 \sum_y S^{(0)}(x,y)\delta D_{\rm b}^{(2)} H^{(0)}(y), \tag{4.29}$$

which arises from the O($a$) boundary counterterm (2.4). Since

$$\begin{aligned}\delta D_{\rm b}^{(2)}\psi(x) = \tilde{c}_{\rm t}^{(1)}\frac{1}{a}\Big\{&\delta_{x_0,a}\left[\psi(x) - P_+\psi(x-a\hat{0})\right] \\ &+ \delta_{x_0,T-a}\left[\psi(x) - P_-\psi(x+a\hat{0})\right]\Big\},\end{aligned} \tag{4.30}$$

the integration over the gluon field is trivial in this case. Moreover, using the boundary conditions satisfied by the free propagator, it is possible to show that the associated contribution,

$$f_{\rm X}^{(1)}(x_0)_{\rm b} = {\rm Re}\ {\rm tr}\big\{H^{(0)}(x)^\dagger \Gamma_{\rm X} H^{(2)}(x)_{\rm b}\big\}, \tag{4.31}$$

is of order $a$ (as it should be).

To compute the Feynman diagrams one inserts the time-momentum representation of the free quark and gluon propagators as given in sect. 3 and appendix A. One then has to sum over the time coordinates of the vertices and a three-dimensional loop momentum, i.e. there are at most $(T/a)^2(L/a)^3$ terms to be added. Each term involves a few additions and multiplications of complex $4\times 4$ matrices. Up to lattice sizes of say $64\times 32^3$, the computational work needed to evaluate all diagrams for a given choice of the kinematical parameters is then not very large, particularly if the Dirac traces are worked out analytically. To guarantee the correctness of our results we have written two independent sets of programs (one per author).

## 5. Computation of $c_{\mathrm{A}}^{(1)}$

From the one-loop results for $f_{\mathrm{A}}(x_0)$ and $f_{\mathrm{P}}(x_0)$ with $C = C' = 0$ it is possible to extract the leading term in the expansion

$$c_{\mathrm{A}} = c_{\mathrm{A}}^{(1)} g_0^2 + c_{\mathrm{A}}^{(2)} g_0^4 + \ldots \tag{5.1}$$

There are several ways in which this calculation can be done. Essentially we shall follow the method outlined in subsect. 6.5 of ref. [1], but to obtain a number of cross-checks we decided to approach the problem on a broader basis, where one studies the continuum limit and $\mathrm{O}(a)$ improvement of several renormalized correlation functions. We shall then also be able to determine the one-loop coefficient $\tilde{c}_{\mathrm{t}}^{(1)}$.

*5.1 Renormalized correlation functions*

To renormalize the correlation functions $f_{\mathrm{A}}(x_0)$ and $f_{\mathrm{P}}(x_0)$ we first of all have to express the bare parameters $m_0$ and $g_0$ through the renormalized ones. We adopt the minimal subtraction scheme described in subsect. 3.4 of ref. [1] and choose

$$m_{\mathrm{R}} = 0. \tag{5.2}$$

To the order considered the required substitution is then given by

$$m_0 = m_{\mathrm{c}}^{(1)} g_{\mathrm{R}}^2 + \mathrm{O}(g_{\mathrm{R}}^4), \tag{5.3}$$

$$g_0^2 = g_{\mathrm{R}}^2 + \mathrm{O}(g_{\mathrm{R}}^4). \tag{5.4}$$

For the improved theory the critical bare mass has first been worked out by Wohlert [5]. His result,

$$am_{\mathrm{c}}^{(1)} = -0.2025564(4) \times C_{\mathrm{F}}, \qquad C_{\mathrm{F}} = \frac{N^2-1}{2N}, \tag{5.5}$$

has later been reproduced in refs. [8,13] using different methods.

For the renormalization of $f_{\mathrm{A}}(x_0)$ and $f_{\mathrm{P}}(x_0)$ we also need the renormalization constants $Z_{\mathrm{A}}$, $Z_{\mathrm{P}}$ and $Z_\zeta$ to one-loop order. So far the latter has only been computed in the framework of dimensional regularization [12]. Since the renormalization group assumes the same form in all mass-independent renormalization schemes, with coefficients related by finite renormalizations, it is then possible to deduce that

$$Z_{\zeta}^{(1)} = -\frac{3}{16\pi^2} C_{\rm F} \ln(a\mu). \tag{5.6}$$

The renormalization constants of the axial current and density are given by

$$Z_{\rm A}^{(1)} = 0, \tag{5.7}$$

$$Z_{\rm P}^{(1)} = \frac{6}{16\pi^2} C_{\rm F} \ln(a\mu). \tag{5.8}$$

They have been calculated directly in the lattice theory by considering matrix elements between quark states [8,13].

The renormalized O($a$) improved amplitudes $[f_{\rm A}(x_0)]_{\rm R}$ and $[f_{\rm P}(x_0)]_{\rm R}$ are now defined by

$$[f_{\rm A}(x_0)]_{\rm R} = Z_{\rm A} Z_{\zeta}^2 \left\{ f_{\rm A}(x_0) + a f_{\delta{\rm A}}(x_0) \right\}, \tag{5.9}$$

$$[f_{\rm P}(x_0)]_{\rm R} = Z_{\rm P} Z_{\zeta}^2 f_{\rm P}(x_0). \tag{5.10}$$

Note that the factors of the form $1 + b(g_0^2) a m_{\rm q}$ need not be written here since we are at zero renormalized quark mass which is equivalent to setting $m_{\rm q} = 0$. The correlation function $f_{\delta{\rm A}}(x_0)$ is defined in the same way as $f_{\rm A}(x_0)$ with $A_\mu^a(x)$ replaced by the counterterm $\delta A_\mu^a(x)$ [cf. eqs. (I.2.26) and (I.6.1)]. To one-loop order we have

$$[f_{\rm A}(x_0)]_{\rm R} = f_{\rm A}^{(0)}(x_0) + g_{\rm R}^2 \Bigg\{ f_{\rm A}^{(1)}(x_0) + m_{\rm c}^{(1)} \frac{\partial}{\partial m_0} f_{\rm A}^{(0)}(x_0) + \left( Z_{\rm A}^{(1)} + 2 Z_{\zeta}^{(1)} \right) f_{\rm A}^{(0)}(x_0) + a f_{\delta{\rm A}}^{(1)}(x_0) \Bigg\} + {\rm O}(g_{\rm R}^4), \tag{5.11}$$

$$[f_{\rm P}(x_0)]_{\rm R} = f_{\rm P}^{(0)}(x_0) + g_{\rm R}^2 \Bigg\{ f_{\rm P}^{(1)}(x_0) + m_{\rm c}^{(1)} \frac{\partial}{\partial m_0} f_{\rm P}^{(0)}(x_0) + \left( Z_{\rm P}^{(1)} + 2 Z_{\zeta}^{(1)} \right) f_{\rm P}^{(0)}(x_0) \Bigg\} + {\rm O}(g_{\rm R}^4), \tag{5.12}$$

where all amplitudes on the right hand sides are to be evaluated at $m_0 = 0$. Recall that

$$f_{\rm X}^{(1)}(x_0) = f_{\rm X}^{(1)}(x_0)_{\rm d} + f_{\rm X}^{(1)}(x_0)_{\rm b} \tag{5.13}$$

with $f_{\rm X}^{(1)}(x_0)_{\rm d}$ being the sum of the diagrams no. 1–3 and $f_{\rm X}^{(1)}(x_0)_{\rm b}$ the contribution of the boundary counterterm proportional to $\tilde{c}_{\rm t}^{(1)}$. The explicit $\mathrm{O}(a)$ counterterm appearing in eq. (5.11) is given by

$$f_{\delta\rm A}^{(1)}(x_0) = c_{\rm A}^{(1)}\tfrac{1}{2}(\partial_0^* + \partial_0)f_{\rm P}^{(0)}(x_0). \tag{5.14}$$

The renormalized amplitudes are thus numerically known and we may proceed to study their behaviour in the continuum limit.

In the following we set

$$T = 2L, \qquad \theta_k = \theta \geq 0, \qquad \mu = 1/L, \tag{5.15}$$

and consider the dimensionless functions

$$h_{\rm P}(\theta, a/L) = [f_{\rm P}(x_0)]_{\rm R}\big|_{x_0=T/2}, \tag{5.16}$$

$$h_{\rm A}(\theta, a/L) = [f_{\rm A}(x_0)]_{\rm R}\big|_{x_0=T/2}, \tag{5.17}$$

$$h_{\rm dA}(\theta, a/L) = L\tfrac{1}{2}(\partial_0^* + \partial_0)[f_{\rm A}(x_0)]_{\rm R}\big|_{x_0=T/2}. \tag{5.18}$$

With properly adjusted coefficients of the $\mathrm{O}(a)$ counterterms all these amplitudes are expected to converge to the continuum limit with a rate proportional to $a^2$. To the order of perturbation theory considered, we only require the value of $c_{\rm sw}$ at $g_0 = 0$ which has long been shown to be equal to 1 [4]. The coefficients to be determined then are $c_{\rm A}^{(1)}$ and $\tilde{c}_{\rm t}^{(1)}$.

Taking eqs. (5.11)–(5.14) into account the perturbation expansion of the $h_{\rm P}$, $h_{\rm A}$ and $h_{\rm dA}$ is written in the form

$$h_{\rm P} = u_0 + g_{\rm R}^2\Big\{u_1 + \tilde{c}_{\rm t}^{(1)}u_2 + am_{\rm c}^{(1)}u_3 + \left(Z_{\rm P}^{(1)} + 2Z_\zeta^{(1)}\right)u_0\Big\} + \mathrm{O}(g_{\rm R}^4), \tag{5.19}$$

$$h_{\rm A} = v_0 + g_{\rm R}^2\Big\{v_1 + \tilde{c}_{\rm t}^{(1)}v_2 + am_{\rm c}^{(1)}v_3 + \left(Z_{\rm A}^{(1)} + 2Z_\zeta^{(1)}\right)v_0 + c_{\rm A}^{(1)}v_4\Big\} + \mathrm{O}(g_{\rm R}^4), \tag{5.20}$$

$$h_{\mathrm{dA}} = w_0 + g_{\mathrm{R}}^2 \Big\{ w_1 + \tilde{c}_{\mathrm{t}}^{(1)} w_2 + am_{\mathrm{c}}^{(1)} w_3 +$$

$$\left( Z_{\mathrm{A}}^{(1)} + 2Z_{\zeta}^{(1)} \right) w_0 + c_{\mathrm{A}}^{(1)} w_4 \Big\} + \mathrm{O}(g_{\mathrm{R}}^4). \qquad (5.21)$$

With the exception of $u_1$, $v_1$ and $w_1$, all coefficients appearing in these equations can be calculated analytically using the results of sect. 3. In the present context we only need to work out their asymptotic form at small $a/L$. Explicit expressions, valid up to corrections of order $(a/L)^2$, are listed in appendix B.

*5.2 Computation of $m_{\mathrm{c}}^{(1)}$*

Since the renormalized quark mass has been set to zero, the axial current is conserved (up to corrections of order $a^2$) and we conclude that

$$h_{\mathrm{dA}}(\theta, a/L) = \mathrm{O}(a^2). \qquad (5.22)$$

In particular, the curly bracket in eq. (5.21) is of order $a^2$. The same is true for the coefficients $w_0$ and $w_2$, while $w_3$ grows proportionally to $L/a$ for $a \to 0$ (cf. appendix B). It follows from this that

$$am_{\mathrm{c}}^{(1)} = -(w_1 + c_{\mathrm{A}}^{(1)} w_4)/w_3 + \mathrm{O}(a^3). \qquad (5.23)$$

The term proportional to $c_{\mathrm{A}}^{(1)}$ makes a contribution of order $a^2$. The critical bare mass may hence be computed by extrapolating the ratio $-w_1/w_3$ to the continuum limit. The extrapolation is particularly easy if $\theta = 0$, because $w_4$ is of order $a^2$ in this case. As a consequence we have

$$am_{\mathrm{c}}^{(1)} = -(w_1/w_3)_{\theta=0} + \mathrm{O}(a^3), \qquad (5.24)$$

i.e. the rate of convergence is improved from order $a^2$ to order $a^3$.

The only dependence of $h_{\mathrm{dA}}$ on the lattice spacing is through the combination $a/L$. Taking the continuum limit thus amounts to calculating the coefficients $w_k$ for a range of lattice sizes $L/a$ and extrapolating to $L/a = \infty$. On general grounds one expects that an asymptotic expansion of the form

$$w_1/w_3 = a_0 + \sum_{k=1}^{\infty} [a_k + b_k \ln(a/L)](a/L)^k \qquad (5.25)$$

holds [14]. The extrapolation method described in sect. 6 of ref. [14], adapted to series with both even and odd powers of $a/L$, may thus be applied.

The coefficients $w_k$ (and also $u_k$ and $v_k$) have been calculated for $L/a = 4, 5, \ldots, 32$ and three values of $\theta$, equal to 0.0, 0.1 and 1.0. The extrapolation of the data at $\theta = 0.0$ then yields

$$am_{\rm c}^{(1)} = -0.2025565(1) \times C_{\rm F}. \tag{5.26}$$

This result is consistent with and slightly more accurate than Wohlert's number eq. (5.5). Consistent results are also obtained at the other values of $\theta$.

*5.3 Logarithmically divergent terms*

The additive mass renormalization constant determined above cancels the linearly divergent terms in the bare one-loop amplitudes $u_1$, $v_1$ and $w_1$. The terms proportional to $\ln(a/L)$ are cancelled by the renormalization constants $Z_{\rm A}^{(1)}$, $Z_{\rm P}^{(1)}$ and $Z_{\zeta}^{(1)}$ given in subsect. 5.1. From the data for $u_1$ and $v_1$ we can now check that the logarithmically divergent terms are indeed as expected.

From eq. (5.19) we infer that

$$Z_{\rm P}^{(1)} + 2Z_{\zeta}^{(1)} = \text{constant} - (u_1 + \tilde{c}_{\rm t}^{(1)} u_2 + am_{\rm c}^{(1)} u_3)/u_0 + {\rm O}(a^2). \tag{5.27}$$

The term proportional to $\tilde{c}_{\rm t}^{(1)}$ makes a contribution proportional to $a/L$ to the right hand side of this equation. Taking eq. (5.24) into account we obtain

$$Z_{\rm P}^{(1)} + 2Z_{\zeta}^{(1)} = \text{constant} - (u_1 - (w_1/w_3)_{\theta=0} u_3)/u_0 + {\rm O}(a). \tag{5.28}$$

It also follows from the above that the error term is reduced to ${\rm O}(a^2)$ if $\theta = 0$.

Before extrapolating the right hand side of eq. (5.28) to the continuum limit the constant must be removed by "differentiating" with respect to $L$, i.e. one takes the difference of the elements of the series at $L+a$ and $L-a$ and multiplies with $-L/2a$. In this way one obtains a series which extrapolates to the coefficient of the $\ln(a/L)$ term. The extrapolation then shows that the series is compatible with zero as the asymptotic value. A conservative estimate of a possible non-zero value is

$$Z_{\rm P}^{(1)} + 2Z_{\zeta}^{(1)} = \frac{c}{16\pi^2} C_{\rm F} \ln(a\mu), \qquad |c| \leq 0.002, \tag{5.29}$$

thus verifying the expected result to high precision.

In the same way it is possible to extract $Z_{\mathrm{A}}^{(1)} + 2Z_{\zeta}^{(1)}$ from the one-loop contribution to $h_{\mathrm{A}}$. Again it is advantageous to set $\theta = 0$ and the result

$$Z_{\mathrm{A}}^{(1)} + 2Z_{\zeta}^{(1)} = \frac{\tilde{c}}{16\pi^2} C_{\mathrm{F}} \ln(a\mu), \qquad \tilde{c} = -5.9999(1), \tag{5.30}$$

is then obtained in complete agreement with eqs. (5.6),(5.7).

*5.4 Computation of* $c_{\mathrm{A}}^{(1)}$

Since $h_{\mathrm{dA}}$, $w_0$ and $w_2$ are all of order $a^2$ it follows that

$$c_{\mathrm{A}}^{(1)} = [-w_1 + (w_1/w_3)_{\theta=0} w_3]/w_4 + \mathrm{O}(a), \tag{5.31}$$

where $\theta$ should be taken non-zero ($w_4$ vanishes otherwise). The extrapolation of the data at $\theta = 0.1$ then yields

$$c_{\mathrm{A}}^{(1)} = -0.00567(1) \times C_{\mathrm{F}}. \tag{5.32}$$

The data at $\theta = 1.0$ can also be extrapolated with consistent results. It should be noted, however, that in this case the order $a^2$ corrections (which here appear as $a/L$ corrections) are almost as large as the leading term. A careful extrapolation procedure is then essential to obtain the correct asymptotic value of the series.

It may not be totally obvious that the calculation of $c_{\mathrm{A}}^{(1)}$ presented here is in line with the computational strategy discussed in subsect. 6.5 of ref. [1]. The proposition made there was to determine $c_{\mathrm{A}}$ from the requirement that the unrenormalized current quark mass $m$, defined through eq. (I.6.13), should be independent of the kinematical parameters. Since the renormalized quark mass has been set to zero, the criterion is that $m$ must vanish up to $\mathrm{O}(a^2)$ corrections. From the definitions of $m$ and $h_{\mathrm{dA}}$ it is straightforward to show that

$$h_{\mathrm{dA}} = Z_{\mathrm{A}} Z_{\zeta}^2 L \left\{ 2m f_{\mathrm{P}}(x_0) + \tfrac{1}{4} c_{\mathrm{A}} a^3 (\partial_0^* \partial_0)^2 f_{\mathrm{P}}(x_0) \right\}_{x_0 = T/2}. \tag{5.33}$$

The last term in this equation is manifestly of order $a^3$ so that $h_{\mathrm{dA}}$ and $m$ are practically proportional to each other. Requiring $m$ to be of order $a^2$ or $h_{\mathrm{dA}}$ (as we did here) are hence equivalent conditions.

*5.5 Computation of $\tilde{c}_{\rm t}^{(1)}$*

To compute $\tilde{c}_{\rm t}^{(1)}$ the best quantity to consider appears to be the ratio

$$\frac{h_{\rm P}(\theta, a/L)}{h_{\rm P}(0, a/L)}. \tag{5.34}$$

The requirement of O($a$) improvement at one-loop order then implies that

$$\begin{aligned}\tilde{c}_{\rm t}^{(1)} u_2/u_0 &= \text{constant} - u_1/u_0 + (u_1/u_0)_{\theta=0} \\ &\quad - a m_{\rm c}^{(1)} \left[u_3/u_0 - (u_3/u_0)_{\theta=0}\right] + {\rm O}(a^2). \end{aligned} \tag{5.35}$$

Taking the difference of this equation evaluated at $L+a$ and $L-a$ and multiplying with the appropriate factor then results in a series which extrapolates to $\tilde{c}_{\rm t}^{(1)}$. Again it is found that the O($a/L$) corrections in this series are not small. For both values, $\theta = 0.1$ and $\theta = 1.0$, the extrapolation yields

$$\tilde{c}_{\rm t}^{(1)} = -0.0135(1) \times C_{\rm F}. \tag{5.36}$$

As a further check on this result one may verify that the same number is obtained from a similar analysis of the ratio

$$\frac{h_{\rm A}(\theta, a/L)}{h_{\rm A}(0, a/L)}, \tag{5.37}$$

where the value (5.32) is inserted for $c_{\rm A}^{(1)}$.

## 6. Non-zero background fields and computation of $c_{\rm sw}^{(1)}$

In this section we outline the computation of the first two coefficients in the expansion

$$c_{\rm sw} = c_{\rm sw}^{(0)} + c_{\rm sw}^{(1)} g_0^2 + c_{\rm sw}^{(2)} g_0^4 + \ldots \tag{6.1}$$

by demanding the O($a$) improvement of the renormalized correlation function $h_{\rm dA}$, defined in subsect. 5.1, for non-zero boundary values $C$ and $C'$. To simplify the calculation we drop all O($a$) boundary counterterms. This is permissible, because the improvement of $h_{\rm dA}$ does not depend on these terms (cf. subsect. 6.1 of ref. [1]).

*6.1 Perturbation theory*

We restrict attention to constant abelian boundary values as specified in subsect. 6.2 of ref. [1]. For the gauge field $V(x,\mu)$ which minimizes the Wilson plaquette action one then obtains [9]

$$V(x,0)=1, \qquad V(x,k)=\exp\{aB_k(x_0)\}, \tag{6.2}$$

$$B_k(x_0)=(C_k'-C_k)x_0/T+C_k. \tag{6.3}$$

This configuration is referred to as the background field in the following.

The perturbation expansion of $h_{\rm dA}$ is derived essentially as before. The main difference is that one expands around the background field, i.e. eq. (4.5) is replaced by

$$U(x,\mu)=V(x,\mu)\exp\{g_0 aq_\mu(x)\}. \tag{6.4}$$

This has already been discussed at length in refs. [9,11] for the case of the pure SU(2) gauge theory. Most of the details given there on the gauge fixing and the construction of the gluon and ghost propagators and vertices can be taken over literally. The extension from SU(2) to SU(3) is straightforward.

As far as the quark fields are concerned we proceed as in sect. 4. In particular, eqs. (4.2),(4.3) are still valid and the expansion of $\psi_{\rm cl}(x)$ in powers of $g_0$ is obtained as in subsect. 4.3. For reasons made clear below we now also need to expand the quark contribution

$$\begin{aligned}\operatorname{tr}\{\ln(D+\delta D+m_0)\}= &\operatorname{tr}\{\ln(D^{(0)}+\delta D^{(0)}+m_0)\}\\ &+g_0\operatorname{tr}\{(D^{(1)}+\delta D^{(1)})(D^{(0)}+\delta D^{(0)}+m_0)^{-1}\}+\ldots\end{aligned} \tag{6.5}$$

to the total action. The operators occurring in this equation act in the space of quark fields with zero boundary values. Note that the first order correction does not vanish for non-trivial background fields.

*6.2 Determination of $c_{\rm sw}^{(0)}$*

Although we already know that $c_{\rm sw}^{(0)}=1$ from the original work of Sheikholeslami and Wohlert [4], it is instructive to rederive this result here by examining the cutoff dependence of $h_{\rm dA}$ at tree-level of perturbation theory. The calculation can be done analytically in spite of the fact that an explicit expression for the quark propagator at $g_0=0$ is not available. We are only interested in evaluating $h_{\rm dA}$ at zero quark mass and thus set $m_0 = 0$ throughout this subsection.

As in the case of vanishing boundary values we have

$$f_{\rm A}^{(0)}(x_0) = -\tfrac{1}{2}\,{\rm tr}\{H^{(0)}(x)^{\dagger}\gamma_0 H^{(0)}(x)\}, \qquad (6.6)$$

where the matrix $H^{(0)}(x)$ is defined through

$$(D^{(0)} + \delta D^{(0)})H^{(0)}(x) = 0, \qquad 0 < x_0 < T, \qquad (6.7)$$

$$P_+ H^{(0)}(x)|_{x_0=0} = P_+, \qquad P_- H^{(0)}(x)|_{x_0=T} = 0. \qquad (6.8)$$

(cf. sects. 2 and 4). The solution of these equations has a relatively simple form as a result of the special properties of the background field. First note that

$$B_k(x_0) = ib(x_0) \qquad (6.9)$$

for all $k = 1, 2, 3$, where $b(x_0)$ is a linear function of $x_0$ with coefficients that are real diagonal matrices in colour space. In particular, the background field is invariant under space translations and the solution $H^{(0)}(x)$ is hence a function of the time $x_0$ only. Eq. (6.7) thus becomes

$$\{P_+\partial_0^* - P_-\partial_0 + \mathcal{A}(x_0) + i\mathcal{B}(x_0)\gamma + i\mathcal{C}(x_0)\gamma_0\gamma\}H^{(0)}(x) = 0, \qquad (6.10)$$

where $\gamma = \gamma_1 + \gamma_2 + \gamma_3$ and

$$\mathcal{A}(x_0) = \frac{6}{a}\left\{\sin\left[\tfrac{1}{2}a(b(x_0) + \theta/L)\right]\right\}^2, \qquad (6.11)$$

$$\mathcal{B}(x_0) = \frac{1}{a}\sin\left[a(b(x_0) + \theta/L)\right], \qquad (6.12)$$

$$\mathcal{C}(x_0) = -\frac{c_{\rm sw}^{(0)}}{2a}\sin\left[a^2\partial_0 b(x_0)\right]. \qquad (6.13)$$

To solve eq. (6.10) we make the ansatz

$$H^{(0)}(x) = \left[s(x_0) + it(x_0)\gamma\right]P_+. \qquad (6.14)$$

The Dirac equation and the boundary conditions then reduce to

$$\partial_0^* s = 3(\mathcal{B} + \mathcal{C})t - \mathcal{A}s, \qquad s(0) = 1, \qquad (6.15)$$

$$\partial_0 t = (\mathcal{B} - \mathcal{C})s + \mathcal{A}t, \qquad t(T) = 0. \tag{6.16}$$

Note that the coefficients $\mathcal{A}$, $\mathcal{B}$, $\mathcal{C}$ and the amplitudes $s$ and $t$ are all diagonal matrices in colour space with real entries. The solution of the equations can be worked out numerically, but this will not be required in the following.

From eqs. (6.6) and (6.14) we now deduce that

$$f_{\rm A}^{(0)}(x_0) = \,{\rm tr}\{-s(x_0)^2 + 3t(x_0)^2\}, \tag{6.17}$$

where the trace is over the colour indices. Using the equations above it is then straightforward to show that

$$\begin{aligned} &\tfrac{1}{2}(\partial_0^* + \partial_0) f_{\rm A}^{(0)}(x_0) = \\ &\qquad {\rm tr}\{c_1(x_0)s(x_0)^2 + c_2(x_0)t(x_0)^2 + c_3(x_0)s(x_0)t(x_0)\} \end{aligned} \tag{6.18}$$

with coefficients $c_1$, $c_2$ and $c_3$ given explicitly in terms of $\mathcal{A}$, $\mathcal{B}$ and $\mathcal{C}$. In particular, their behaviour in the continuum limit is analytically calculable and one finds that $c_1$ and $c_2$ are of order $a^2$ while

$$c_3(x_0) = a(c_{\rm sw}^{(0)} - 1)\partial_0 b(x_0) + {\rm O}(a^2). \tag{6.19}$$

The amplitudes $s$ and $t$ converge to some well-defined smooth functions in the continuum limit and we thus conclude that

$$h_{\rm dA}^{(0)} = a(c_{\rm sw}^{(0)} - 1)L\,{\rm tr}\{\partial_0 b(x_0)s(x_0)t(x_0)\}_{x_0=T/2} + {\rm O}(a^2). \tag{6.20}$$

The trace can be worked out explicitly using the formulae given in appendix C and it is then immediate that it does not vanish for general boundary values of the specified type. For O($a$) improvement the coefficient $c_{\rm sw}^{(0)}$ must hence be set to 1.

### *6.3 Computation of $c_{\rm sw}$ to one-loop order*

The one-loop diagrams contributing to $f_{\rm X}(x_0)$ are listed in figs. 1 and 2. The lines in these diagrams now represent the gluon, quark and ghost propagators in the given background field. The tadpole diagrams in fig. 2 with the quark loop arise from the combination of the first order term in the expansion (6.5) with the first order correction to $\mathrm{tr}\{H(x)^\dagger \Gamma_{\mathrm{X}} H(x)\}$. These are the only diagrams proportional to the number $N_{\mathrm{f}}$ of quarks.

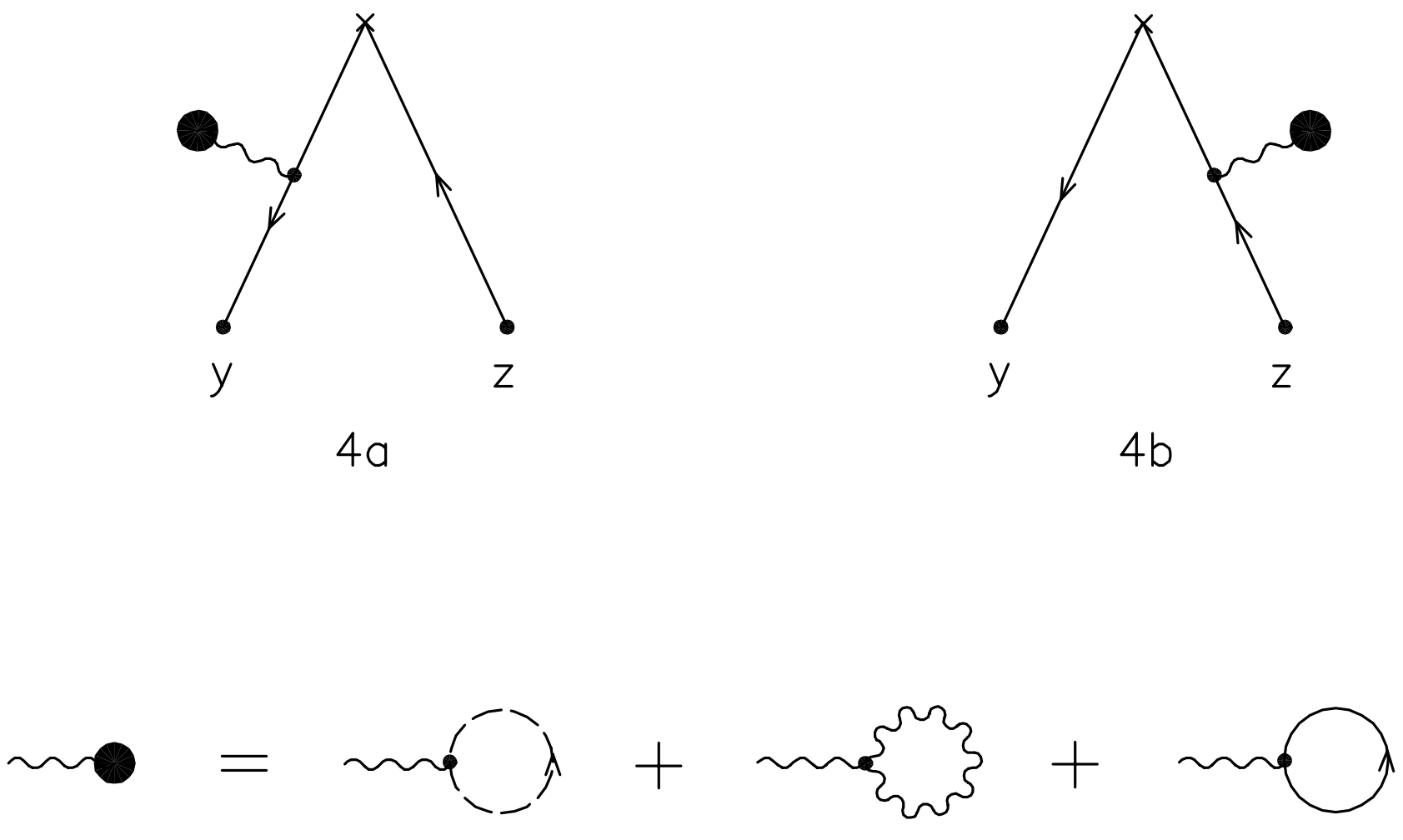

Fig. 2. Tadpole diagrams contributing to $f_{\mathrm{X}}(x_0)$. The directed dashed line represents the ghost propagator.

Since there is no general closed analytic expression for the propagators in the background field (6.2),(6.3), the computer time required for the calculation of the diagrams is not negligible. We have hence decided to set $T = L$ (rather than $T = 2L$) and to produce data only for even $L/a$ up to $L/a = 32$. As for the boundary values of the gauge field we made the same choices as in ref. [10]. We now also set $c_{\mathrm{sw}}^{(0)} = 1$, $\theta = 0$ or $\theta = 1$ and perform all calculations at zero quark mass.

The perturbation expansion of $h_{\mathrm{dA}}$ to order $g_{\mathrm{R}}^2$ assumes the form

$$h_{\mathrm{dA}} = w_0 + g_{\mathrm{R}}^2 \Big\{ w_1 + am_{\mathrm{c}}^{(1)} w_3 + \left( Z_{\mathrm{A}}^{(1)} + 2Z_{\zeta}^{(1)} \right) w_0 \\ + c_{\mathrm{A}}^{(1)} w_4 + c_{\mathrm{sw}}^{(1)} w_5 \Big\} + \mathrm{O}(g_{\mathrm{R}}^4). \qquad (6.21)$$

The coefficients $w_0, w_1, \ldots$ are not the same as in eq. (5.21), but their definitions are completely analogous. Since $w_0$ is of order $a^2$ the corresponding term in the curly bracket may be dropped in the following. From the discussion in subsect. 6.2 one deduces that

$$w_5 = aL\,\mathrm{tr}\{\partial_0 b(x_0)s(x_0)t(x_0)\}_{x_0=T/2} + \mathrm{O}(a^2). \qquad (6.22)$$

For the chosen boundary values of the gauge field, the trace in this expression does not vanish. We thus conclude that

$$c_{\mathrm{sw}}^{(1)} = -(w_1 + am_{\mathrm{c}}^{(1)} w_3 + c_{\mathrm{A}}^{(1)} w_4)/w_5 + \mathrm{O}(a/L). \qquad (6.23)$$

Using the value of $c_{\mathrm{A}}^{(1)}$ given in eq. (5.32), the extrapolation of our data then yields

$$c_{\mathrm{sw}}^{(1)} = \begin{cases} 0.155(1) & \text{for } N = 2, \\ 0.267(1) & \text{for } N = 3. \end{cases} \qquad (6.24)$$

There is no dependence on $N_{\mathrm{f}}$, because the contribution to $w_1$ of the quark tadpole diagrams turns out to be of order $a^2$.

Our computations have been checked in various ways. For $N = 2$ the expressions for the ghost and gluon tadpoles can be compared with the corresponding expressions given in ref. [11]. We have also verified that the total one-loop contributions to $f_{\mathrm{A}}(x_0)$ and $f_{\mathrm{P}}(x_0)$ are independent of the gauge parameter. For this the gluon tadpole diagrams are essential. In addition runs up to $L/a = 20$ were made with alternative choices of the boundary values $C$ and $C'$. In all cases the values quoted above for $c_{\mathrm{sw}}^{(1)}$ have been reproduced, although to a reduced precision. Finally by combining our data at $\theta = 0$ and $\theta = 1$, it is possible to extract the coefficient $c_{\mathrm{A}}^{(1)}$. The values that have been obtained in this way are less accurate but completely consistent with eq. (5.32).

Our results confirm those obtained by Wohlert [5] nearly 10 years ago, who actually cites numbers with a higher accuracy, viz.

$$c_{\mathrm{sw}}^{(1)} = \begin{cases} 0.15400(4) & \text{for } N = 2, \\ 0.26590(7) & \text{for } N = 3. \end{cases} \qquad (6.25)$$

We are unfortunately unable to directly check whether Wohlert's error analysis is realistic since his raw data are no longer available. But our result, eq.(6.24), is perfectly satisfactory for application at small bare coupling. At higher values of $g_0$ one should in any case compute $c_{\mathrm{sw}}$ non-perturbatively [2].

Table 1. Values of the coefficients $r_0$ and $r_1$ at $m_0 = m_c$, $T = 2L$ and $x_0 = T/2$

| $\theta$ | $L/a$ | $r_0$ | $r_1$ | $r_1\vert_{c_A^{(1)}=0}$ |
|---|---|---|---|---|
| 0.0 | 6 | 0.0 | 0.000861 | 0.000861 |
| | 8 | 0.0 | 0.000175 | 0.000175 |
| | 10 | 0.0 | 0.000055 | 0.000055 |
| | 12 | 0.0 | 0.000023 | 0.000023 |
| | 14 | 0.0 | 0.000011 | 0.000011 |
| | 16 | 0.0 | 0.000006 | 0.000006 |
| 1.0 | 6 | 0.001151 | 0.001027 | 0.001963 |
| | 8 | 0.000365 | 0.000254 | 0.000782 |
| | 10 | 0.000150 | 0.000092 | 0.000431 |
| | 12 | 0.000072 | 0.000042 | 0.000278 |
| | 14 | 0.000039 | 0.000022 | 0.000195 |
| | 16 | 0.000023 | 0.000013 | 0.000146 |

## 7. How large are the remaining cutoff effects?

To answer this question we here consider a few quantities constructed from $f_A$ and $f_P$ and study their approach to the continuum limit. For simplicity we restrict attention to the case of vanishing boundary values of the gauge field (the situation is not very different at non-zero values of $C$ and $C'$).

### *7.1 PCAC relation*

As in sect. 5 we again set $T = 2L$, $\theta_k = \theta$ and study the theory at the critical point $m_0 = m_c$. The unrenormalized current quark mass $m$, defined through eq. (I.6.13), is then expected to vanish up to terms of order $a^2$. The coefficients $r_0$ and $r_1$ in the expansion

$$am = r_0 + r_1 C_F g_0^2 + O(g_0^4) \qquad (7.1)$$

should hence go to zero in the continuum limit with a rate roughly proportional to $(a/L)^3$. The numbers listed in table 1 show that at $x_0 = T/2$ both, $r_0$ and $r_1$, are indeed small and rapidly approaching zero.

It is interesting to note in this connection that the contribution of the $O(a)$ correction proportional to $c_A^{(1)}$ is negligible at small values of $\theta$, while at $\theta = 1$ a visible improvement is achieved by including this term. The contribution of the boundary counterterm proportional to $\tilde{c}_t^{(1)}$, on the other hand, appears to be zero within rounding errors and is certainly completely negligible.

The values of $am$ that one obtains from eq. (7.1) and the coefficients listed in table 1 are quite small. At $L/a = 8$, $\theta = 1$, $g_0^2 = 1$ and $N = 3$, for example, the value $am = 0.0007$ is found. In the quenched approximation, where the cutoff $1/a$ is about 2 GeV at $g_0^2 = 1$, this translates to a mass $m \simeq 1.4$ MeV. As discussed in subsect. 6.6 of ref. [1], this may be taken as an indication for the systematic uncertainty which one has when defining the zero mass point on a $16 \times 8^3$ lattice. By increasing the lattice size to say $L/a = 12$ (which is still rather small by normal standards) the uncertainty is reduced by a factor 6 at $\theta = 1$ and nearly an order of magnitude at $\theta = 0$. One should however be warned that perturbation theory can only give a rough idea on the actual size of the cutoff effects. In particular, at the larger couplings the lattice effects associated with the non-perturbative scales may become important (cf. ref. [2]).

*7.2 Another test of PCAC*

So far we have set $x_0 = T/2$ in the definition of $m$, but up to $O(a^2)$ corrections the same value of $m$ should be obtained for all $x_0$. As discussed in sect. 3 the lowest order coefficient $r_0$ is in fact independent of $x_0$.

The behaviour of $r_1$ as a function of $x_0$ on a $16 \times 8^3$ lattice is shown in fig. 3. A significant increase of the cutoff effects towards the boundaries of the lattice is observed, but it should also be said that they remain small and certainly are not above the expected order of magnitude. The fact that the effects are larger near the boundaries may have several causes. An obvious possibility is that the contributions of the high-energy states to the correlation functions $f_A(x_0)$ and $f_P(x_0)$ are not sufficiently suppressed if $x_0$ or $T - x_0$ is small in lattice units. But it could also be that the effects seen in fig. 3 arise from the $O(a^2)$ corrections associated with the low-energy intermediate states.

On larger lattices the cutoff effects are rapidly becoming small and when $r_1$ is plotted against $x_0$ the picture is essentially a scaled version of fig. 3. An interesting detail is that $r_1$ is practically independent of $\theta$ in the middle of the lattice. For this to come out it has been important to include the $O(a)$ correction proportional to $c_A^{(1)}$, which is independent of $x_0$ and approximately given by $6c_A^{(1)}(a\theta/L)^2$.

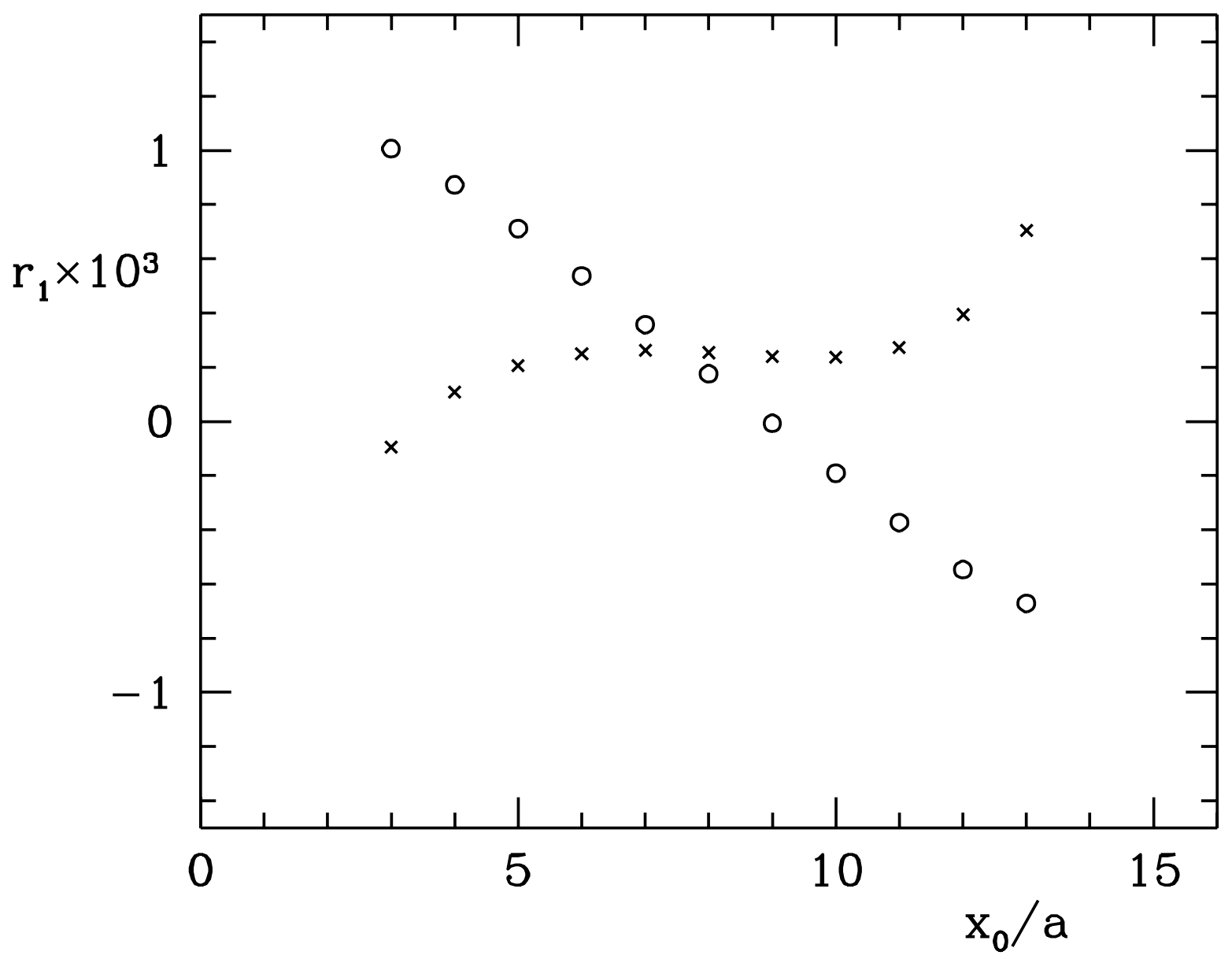

Fig. 3. Values of $r_1$ on a $16 \times 8^3$ lattice as a function of the time $x_0$ at which the axial current is inserted. Circles and crosses are for $\theta = 0$ and $\theta = 1$ respectively.

*7.3 Cutoff dependence of $f_{\rm P}(x_0)$*

The definition of the renormalized correlation function $[f_{\rm P}(x_0)]_{\rm R}$ involves some arbitrariness in the choice of normalization. It may then not be easy to decide whether the cutoff effects that one finds should be considered large or small. More sensible quantities to study are ratios of correlation functions where the normalization factors cancel.

To illustrate this we introduce the function

$$F(\theta, a/L, g_0) = f_{\rm P}(x_0)|_{x_0=T/2}, \tag{7.2}$$

where $T = 2L$, $m_0 = m_{\rm c}$ and $\theta_k = \theta$ as before. The ratio

$$\rho = \frac{F(\theta, a/L', g_0)}{F(\theta, a/L, g_0)}, \qquad L' = 2L, \tag{7.3}$$

then is a renormalization group invariant, i.e. in the continuum limit $\rho$ converges to a function of $\theta$ and the running coupling $\bar{g}^2$ at scale $L$ (in any scheme).

Table 2. Values of the coefficients $\rho_0$ and $\rho_1$ at $m_0 = m_c$, $T = 2L$ and $x_0 = T/2$

| $\theta$ | $L/a$ | $\rho_0$ | $\rho_1$ | $\rho_1\|_{\tilde{c}_t^{(1)}=0}$ |
|---|---|---|---|---|
| 0.0 | 6 | 1.0 | 0.0236 | 0.0236 |
| | 8 | 1.0 | 0.0101 | 0.0101 |
| | 10 | 1.0 | 0.0056 | 0.0056 |
| | 12 | 1.0 | 0.0036 | 0.0036 |
| | 14 | 1.0 | 0.0025 | 0.0025 |
| | 16 | 1.0 | 0.0019 | 0.0019 |
| 1.0 | 6 | 0.9584 | −0.0049 | −0.0099 |
| | 8 | 0.9755 | −0.0033 | −0.0076 |
| | 10 | 0.9839 | −0.0022 | −0.0059 |
| | 12 | 0.9886 | −0.0016 | −0.0048 |
| | 14 | 0.9916 | −0.0011 | −0.0040 |
| | 16 | 0.9935 | −0.0008 | −0.0034 |

The approach to the continuum limit of the ratio (7.3) can now be studied in perturbation theory. In table 2 the coefficients $\rho_0$ and $\rho_1$ appearing in the expansion

$$\rho = \rho_0 + \rho_1 C_F \bar{g}^2 + O(\bar{g}^4) \tag{7.4}$$

are listed for two values of $\theta$ and a range of $L/a$. In the continuum theory the coefficients are 1 and 0, respectively, for all $\theta$. The cutoff effects thus appear to be rather small and decrease roughly proportionally to $(a/L)^2$ as expected. From the third column in the table we can also see that the $O(a)$ counterterm proportional to $\tilde{c}_t^{(1)}$ is completely negligible at $\theta = 0$. When $\theta$ increases the contribution of the boundary counterterm can become numerically significant compared to the cutoff effects at tree-level, but it remains rather small in absolute terms if (say) $C_F \bar{g}^2 \leq 1$.

## Appendix A

The Fourier representation of the gluon propagator (4.17) reads

$$D_{\mu\nu}(x,y) = L^{-3} \sum_{\mathbf{p}} e^{i\mathbf{p}(\mathbf{x}-\mathbf{y})} d_{\mu\nu}(x_0, y_0; \mathbf{p}). \tag{A.1}$$

For any momentum $\mathbf{p}$ we define the "energy" $\varepsilon \geq 0$ through

$$\cosh(a\varepsilon) = 1 + \tfrac{1}{2} a^2 \hat{\mathbf{p}}^2. \tag{A.2}$$

In view of the symmetry

$$d_{\mu\nu}(x_0, y_0; \mathbf{p}) = d_{\nu\mu}(y_0, x_0; \mathbf{p}), \tag{A.3}$$

it suffices to consider the case $x_0 \geq y_0$. In the Feynman gauge, $\lambda_0 = 1$, and for non-zero momenta $\mathbf{p}$ the propagator is then given by

$$\begin{aligned} d_{00}(x_0, y_0; \mathbf{p}) = & \frac{a}{\sinh(\varepsilon a)\sinh(\varepsilon T)} \\ & \times \cosh[\varepsilon(T - x_0 - \tfrac{1}{2}a)] \cosh[\varepsilon(y_0 + \tfrac{1}{2}a)], \end{aligned} \tag{A.4}$$

$$\begin{aligned} d_{kj}(x_0, y_0; \mathbf{p}) = & \delta_{kj} \frac{a}{\sinh(\varepsilon a)\sinh(\varepsilon T)} \\ & \times \sinh[\varepsilon(T - x_0)] \sinh(\varepsilon y_0), \end{aligned} \tag{A.5}$$

while for $\mathbf{p} = 0$ we have

$$d_{00}(x_0, y_0; \mathbf{0}) = y_0 + a, \tag{A.6}$$

$$d_{kj}(x_0, y_0; \mathbf{0}) = \delta_{kj}(T - x_0)\frac{y_0}{T}. \tag{A.7}$$

The mixed components $d_{0k}$ and $d_{k0}$ vanish for all time coordinates and all momenta $\mathbf{p}$.

## Appendix B

Most of the coefficients appearing in eqs. (5.19)–(5.21) are known analytically and can be expanded in powers of $a/L$. It is convenient to introduce the abbreviations

$$\mathrm{si} = \sinh(2\sqrt{3}\theta), \tag{B.1}$$

$$\mathrm{co} = \cosh(2\sqrt{3}\theta). \tag{B.2}$$

Up to corrections of order $(a/L)^2$ the analytically calculable coefficients are then given by

$$u_0 = \frac{N}{\mathrm{co}}, \tag{B.3}$$

$$u_2 = \frac{12N\theta\mathrm{si}}{\sqrt{3}\mathrm{co}^2}\,\frac{a}{L}, \tag{B.4}$$

$$u_3 = -\frac{N\mathrm{si}}{\sqrt{3}\theta\,\mathrm{co}^2}\,\frac{L}{a} + \left\{\frac{4N\theta^2\left(1-\mathrm{si}^2\right)}{3\,\mathrm{co}^3} - \frac{N\theta\mathrm{si}}{6\sqrt{3}\mathrm{co}^2}\right\}\frac{a}{L}, \tag{B.5}$$

$$v_0 = -\frac{N}{\mathrm{co}^2}, \tag{B.6}$$

$$v_2 = -\frac{24N\theta\mathrm{si}}{\sqrt{3}\mathrm{co}^3}\,\frac{a}{L}, \tag{B.7}$$

$$v_3 = -\frac{N\mathrm{si}\,(\mathrm{co}-2)}{\sqrt{3}\theta\,\mathrm{co}^3}\,\frac{L}{a} - \left\{\frac{N\theta\mathrm{si}\,(\mathrm{co}-2)}{6\sqrt{3}\mathrm{co}^3} + \frac{4N\theta^2}{3\,\mathrm{co}^4}(\mathrm{co}^3 - 4\,\mathrm{co}^2 - 2\,\mathrm{co} + 6)\right\}\frac{a}{L}, \tag{B.8}$$

$$v_4 = -\frac{6N\theta\mathrm{si}}{\sqrt{3}\mathrm{co}^2}\,\frac{a}{L}, \tag{B.9}$$

$$w_0 = 0, \tag{B.10}$$

$$w_2 = 0, \tag{B.11}$$

$$w_3 = \frac{2N}{\text{co}}\,\frac{L}{a} + \left\{\frac{8N\theta^3\,\text{si}}{\sqrt{3}\,\text{co}^2} + \frac{3N\theta^2}{\text{co}}\right\}\frac{a}{L}, \tag{B.12}$$

$$w_4 = \frac{12N\theta^2}{\text{co}}\,\frac{a}{L}. \tag{B.13}$$

## Appendix C

The functions $s$ and $t$ introduced in subsect. 6.2 can be computed analytically in the continuum limit. The equations to be solved are

$$\partial_0 s = 3(b+\theta/L)t, \qquad s(0) = 1, \tag{C.1}$$

$$\partial_0 t = (b+\theta/L)s, \qquad t(T) = 0. \tag{C.2}$$

To be able to write down the solution in a compact form it is useful to define

$$z(x_0) = \frac{\sqrt{3}}{\partial_0 b(x_0)}\left[b(x_0)+\theta/L\right]^2. \tag{C.3}$$

Note that the denominator $\partial_0 b(x_0)$ is independent of $x_0$. One may then easily verify that

$$s(x_0) = k\sqrt{3}\cosh\tfrac{1}{2}\left[z(x_0)-z(T)\right], \tag{C.4}$$

$$t(x_0) = k\sinh\tfrac{1}{2}\left[z(x_0)-z(T)\right], \tag{C.5}$$

solves the system (C.1),(C.2) if we choose

$$k = \left\{\sqrt{3}\cosh\tfrac{1}{2}\left[z(0)-z(T)\right]\right\}^{-1}. \tag{C.6}$$

It is also not difficult to show that the solution is unique.